\documentclass[10pt,draft]{dis03}
\usepackage{epsf,amsmath}

\def\beq{\begin{equation}}
\def\eeq{\end{equation}}
\def\beqa{\begin{eqnarray}}
\def\eeqa{\end{eqnarray}}

\begin{document}

\title{NNLO SOFT AND VIRTUAL CORRECTIONS FOR ELECTROWEAK,
HIGGS, AND SUSY PROCESSES
\thanks{The author's research has been supported 
by a Marie Curie Fellowship 
of the European Community programme ``Improving Human Research Potential'' 
under contract number HPMF-CT-2001-01221.}}

\author{NIKOLAOS KIDONAKIS \\
Cavendish Laboratory, University of Cambridge\\
Cambridge CB3 0HE, England\\
E-mail: kidonaki@hep.phy.cam.ac.uk\\ }

\maketitle

\begin{abstract}
\noindent I present applications of a master formula for 
next-to-next-to-leading order soft and virtual QCD corrections 
to various electroweak, Higgs, and supersymmetric processes.
They include Drell-Yan and charged Higgs production, single-top production in
flavor-changing neutral-current processes, and squark and gluino production.

\end{abstract}

\section{Introduction} 

Soft and virtual QCD corrections to processes of electroweak
or supersymmetric (SUSY) origin can be substantial. The
calculations of the cross sections, total or differential, 
in hadron-hadron and lepton-hadron colliders can be represented 
in factorized form by
\beq
\sigma=\sum_f \int \left[ \prod_i  dx_i \, \phi_{f/h_i}(x_i,\mu_F^2)\right]\,
{\hat \sigma}(s,t_i,\mu_F,\mu_R)
\label{factcs}
\eeq
with $\sigma$ the physical cross section, ${\hat \sigma}$ the
partonic cross section,
$\phi_{f/h_i}$ the parton distribution for parton $f$ in hadron
$h_i$, and $\mu_F$, $\mu_R$ the factorization and renormalization scales, 
respectively.

The perturbatively calculable ${\hat \sigma}$ includes
soft and virtual corrections from soft-gluon
emission and virtual diagrams. These corrections appear as
plus distributions and delta functions in ${\hat \sigma}$.
In single-particle-inclusive (1PI) kinematics the plus distributions are 
${\cal D}_l(s_4)\equiv[\ln^l(s_4/M^2)/s_4]_+$ with $s_4=s+t+u-\sum m^2$,
where $s,t,u$ are kinematical invariants and $m$ the masses of the 
particles in the scattering, and $M$ any relevant hard scale.
In pair-invariant-mass (PIM) kinematics they are
${\cal D}_l(z)\equiv[\ln^l(1-z)/(1-z)]_+$ with $z=Q^2/s$, where
$Q^2$ is the pair mass squared. Note that $s_4$ (sometimes called $s_2$) 
$\rightarrow 0$ and $z$ (sometimes called $x$) $\rightarrow 1$
at threshold.

A unified approach and a master formula for calculating
these corrections at next-to-next-to-leading order (NNLO) for any process in
hadron-hadron and lepton-hadron colliders have been recently presented
in Ref. \cite{NKNNLO}; they follow from threshold resummation studies 
\cite{KS,LOS,NK,2nloop}.
Here I describe various applications to processes which are of electroweak or
supersymmetric origin at lowest order.

\section{NLO and NNLO corrections}

I begin by presenting the generalized next-to-leading-order
(NLO) master formula for processes with
simple color flows, which is appropriate for many electroweak
and SUSY processes.
The NLO soft and virtual corrections in the $\overline{\rm MS}$ scheme 
in either 1PI or PIM kinematics take the form
\beq
{\hat{\sigma}}^{(1)} = \sigma^B \frac{\alpha_s(\mu_R^2)}{\pi}
\left\{c_3 \, {\cal D}_1(x_{th}) + c_2 \, {\cal D}_0(x_{th})
+c_1 \, \delta(x_{th})\right\}\, ,
\label{NLO}
\eeq
with $x_{th}$ the threshold variable $s_4$ (in 1PI kinematics)
or $1-z$ (in PIM kinematics), where $\sigma^B$ is the Born term,
$c_3=\sum_i 2C_{f_i}$,
$c_2=T_2-\sum_i C_{f_i} \ln(\mu_F^2/s)$ with
\beq
T_2=2 \, {\rm Re}{\Gamma'}_S^{(1)}- \sum_i \left[C_{f_i}
+2 C_{f_i} \, \delta_K \, \ln\left(\frac{-t_i}{M^2}\right)\right] \, ,
\eeq
and $c_1 =c_1^{\mu} +T_1$, with
\beq
c_1^{\mu}=\sum_i \left[C_{f_i}\, \delta_K \, \ln\left(\frac{-t_i}{M^2}\right)
-\gamma_i^{(1)}\right]\ln\left(\frac{\mu_F^2}{s}\right)
+d_{\alpha_s} \frac{\beta_0}{4} \ln\left(\frac{\mu_R^2}{s}\right) \, .
\eeq
We note that we sum over incoming partons $i$ and the $C_{f_i}$'s are 
color factors, $C_F=4/3$ for quarks and $C_A=3$ for gluons.
Also $\delta_K$ is 0 (1) for PIM (1PI) kinematics.
$\Gamma_S$ are soft anomalous dimensions which describe the color exchange
in the hard scattering, $\gamma_i$ are parton anomalous
dimensions, $\beta_0=(11C_A-2n_f)/3$ is the lowest-order beta function, and 
$d_{\alpha_s}$ equals 0,1,2 if the Born cross section is of order 
$\alpha_s^0,\alpha_s^1,\alpha_s^2$, respectively.
More detais are given in Ref. \cite{NKNNLO}.

At NNLO the $\overline{\rm MS}$ scheme  master formula for the soft 
and virtual corrections is
${\hat\sigma}^{(2)}=
\sigma^B (\alpha_s^2(\mu_R^2)/\pi^2) {\hat{\sigma'}}^{(2)}$ with
\beqa
{\hat{\sigma'}}^{(2)}&=&
\frac{1}{2} c_3^2 \, {\cal D}_3(x_{th})
+\left[\frac{3}{2} c_3 \, c_2
- \frac{\beta_0}{4} c_3\right] {\cal D}_2(x_{th})
\nonumber \\ && \hspace{-25mm}
{}+\left\{c_3 \, c_1 +c_2^2
-\zeta_2 \, c_3^2 -\frac{\beta_0}{2} \, T_2
+\frac{\beta_0}{4} c_3  \ln\left(\frac{\mu_R^2}{s}\right)
+\sum_i C_{f_i} \, K \right\}
{\cal D}_1(x_{th})
\nonumber \\ && \hspace{-25mm}
{}+\left\{c_2 \, c_1 -\zeta_2 \, c_2 \, c_3+\zeta_3 \, c_3^2
-\frac{\beta_0}{2} T_1
+\frac{\beta_0}{4}\, c_2 \ln\left(\frac{\mu_R^2}{s}\right)
+2 \, {\rm Re}{\Gamma'}_S^{(2)}-\sum_i {\nu}_{f_i}^{(2)}
\right.
\nonumber \\ && \hspace{-25mm} \quad \left.
{}+\sum_i C_{f_i} \left[\frac{\beta_0}{8}
\ln^2\left(\frac{\mu_F^2}{s}\right)
-\frac{K}{2}\ln\left(\frac{\mu_F^2}{s}\right)
-K \, \delta_K \, \ln\left(\frac{-t_i}{M^2}\right)\right]\right\}
{\cal D}_0(x_{th})
\nonumber \\ && \hspace{-25mm}
{}+R_{\delta(x_{th})} \, \delta(x_{th}) \, .
\label{NNLO}
\eeqa
More details and extensions of the master formulas to the more general
case of complex color flows are given in Ref. \cite{NKNNLO}.

\section{Applications to electroweak and SUSY processes}

Using the NNLO master formula I have rederived known NNLO results
for Drell-Yan and Higgs production and for $W^+ \gamma $ production,
and I have produced new results for many other processes \cite{NKNNLO}. 
Here I give a few examples.

\subsection{The Drell-Yan process, $q {\bar q} \rightarrow V$}

The NLO corrections are given by Eq. (\ref{NLO}) with $x_{th}=1-x=1-Q^2/s$,
and $c_3=4C_F$,
$c_2=-2C_F \ln(\mu_F^2/Q^2)$,
$c_1=-(3/2) C_F \ln(\mu_F^2/Q^2)+2C_F\zeta_2-4C_F$. 
Using Eq. (\ref{NNLO}) we rederive the NNLO soft and virtual corrections 
in Ref. \cite{DY} and thus also derive previously unknown 
two-loop anomalous dimensions in Eq. (\ref{NNLO}). 
Similar results are given in Ref. \cite{NKNNLO} for $W^+ \gamma$ production, 
$q {\bar q} \rightarrow W^+ \gamma$ \cite{MSN}, 
and related results are derived in \cite{NKNNLO} 
for Standard Model Higgs production, $gg \rightarrow H$.

\subsection{Charged Higgs production, ${\bar b} g \rightarrow H^+ {\bar t}$}
The NLO corrections are given by Eq. (\ref{NLO}) 
and the NNLO corrections by Eq. (\ref{NNLO}), with $x_{th}=s_2
=s+t+u-m_{H^+}^2-m_{\bar t}^2-m_{\bar b}^2$,
and $c_3=2(C_F+C_A)$,
$c_2=C_F[\ln(m_H^4/(sm_t^2))-1-\ln(\mu_F^2/s)]
+C_A[\ln(m_H^4/(t_1 u_1))-\ln(\mu_F^2/s)]$, 
and $c_1^{\mu}=\ln(\mu_F^2/s)
[C_F \ln(-u_1/m_H^2)+C_A\ln(-t_1/m_H^2)
-3C_F/4-\beta_0/4]+(\beta_0/4) \ln(\mu_R^2/s)$.

\subsection{FCNC single-top production, $eu \rightarrow et$}

Here we consider single-top production mediated via flavor-changing
neutral currents (FCNC) through a term in the effective Langrangian
of the form $\kappa_{tq\gamma} e   \bar t \sigma_{\mu\nu} q
F^{\mu\nu}/\Lambda $ \cite{BK}.

\begin{figure}[!thb]
\vspace*{6.0cm}
\begin{center}
\includegraphics{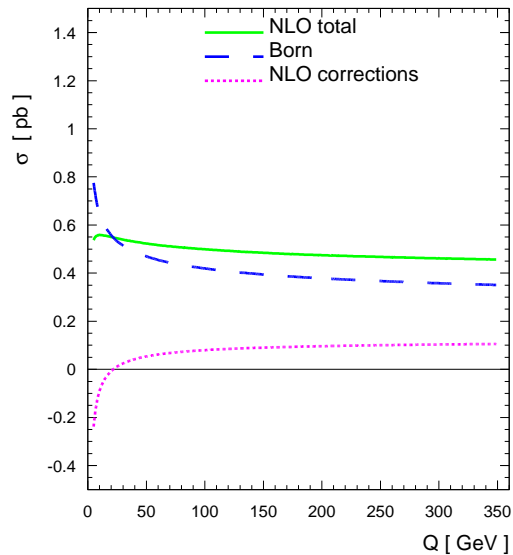}
\vspace*{2.0cm}
\caption[*]{Born cross section, NLO corrections, and total NLO 
(Born+NLO corrections) cross section
for FCNC single-top production at HERA with $m_t$=175 GeV/$c^2$,
$\kappa_{tu\gamma}=0.1$, and $\sqrt{S}=300$ GeV. Here $Q=\mu_F=\mu_R$.}
\end{center}
\end{figure}

The NLO corrections are given by Eq. (\ref{NLO}) with $x_{th}=s_2
=s+t+u-m_t^2-2m_e^2$, $c_3=2C_F$,
$c_2=C_F[-1-2\ln((-u+m_e^2)/m_t^2)
+2\ln((m_t^2-t)/m_t^2)-\ln(\mu_F^2/m_t^2)]$,
and $c_1^{\mu}=[-3/4 +\ln((-u+m_e^2)/m_t^2)]
C_F\ln(\mu_F^2/s)$.
They stabilize the FCNC single top cross section at HERA as a function 
of scale (see fig. 1) \cite{BK}.
The NNLO corrections are given by Eq. (\ref{NNLO}).

\subsection{Squark and gluino production}

We now consider squark and gluino production.
We start with squark pair production.
For the process $q{\bar q} \rightarrow {\tilde q} {\tilde {\bar q}}$
and $qq \rightarrow  {\tilde q} {\tilde q}$
the $c_i$ coefficients are the
same as for the $q{\bar q} \rightarrow Q {\bar Q}$ channel in heavy quark 
pair hadroproduction \cite{NKNNLO};
for $gg \rightarrow {\tilde q} {\tilde {\bar q}}$
they are the same as for $gg \rightarrow Q {\bar Q}$.

We continue with gluino pair production.
For the process
$q{\bar q} \rightarrow {\tilde g} {\tilde g}$ the $c_i$'s are the
same as for $q{\bar q} \rightarrow {\tilde q} {\tilde {\bar q}}$;
for $gg \rightarrow {\tilde g} {\tilde g}$ they are the
same as for $gg \rightarrow {\tilde q} {\tilde {\bar q}}$.

Finally we study squark-gluino production,
$qg \rightarrow {\tilde q} {\tilde g}$.
Here $x_{th}=s_4=s+t+u-m_{\tilde q}-m_{\tilde g}$,
$c_3=2(C_F+C_A)$,
$c_2=-C_F-C_A-2C_F \ln(-u_1/m^2)
-2C_A \ln(-t_1/m^2)-(C_F+C_A)\ln(\mu_F^2/s)$,
and $c_1^{\mu}=\ln(\mu_F^2/s) [C_F\ln(-u_1/m^2)
\newline +C_A \ln(-t_1/m^2)-3C_F/4-\beta_0/4]
+(\beta_0/2) \ln(\mu_R^2/s)$,
with $m=m_{\tilde q}$ or $m_{\tilde g}$.

\end{document}